\documentclass[amssymb,a4style,superscriptaddress]{revtex4}

\usepackage{graphicx}
\usepackage{amsmath}
\usepackage{amsfonts}

\usepackage{bm}% bold math
\newcommand{\ba}{\begin{array}}
\newcommand{\ea}{\end{array}}
\newcommand{\be}{\begin{equation}}
\newcommand{\ee}{\end{equation}}
\newcommand{\bea}{\begin{eqnarray}}
\newcommand{\eea}{\end{eqnarray}}
\font\sqi=cmssq8
\def\DR{\rm I\kern-1.45pt\rm R}
\def\DC{\kern2pt {\hbox{\sqi I}}\kern-4.2pt\rm C}
\def\DH{\rm I\kern-1.5pt\rm H\kern-1.5pt\rm I}

\begin{document}

\title{Cuboctahedric  Higgs oscillator from the rational Calogero model}

\author{Tigran Hakobyan}
\email{hakob@yerphi.am}
\affiliation{Yerevan State University, 1 Alex Manoogian, 375025 Yerevan, Armenia}
\affiliation{Yerevan Physics Institute, 2 Alikhanyan Br., 375036 Yerevan, Armenia}

\author{Armen Nersessian}
\email{arnerses@yerphi.am}
\affiliation{Yerevan State University, 1 Alex Manoogian, 375025 Yerevan, Armenia}
\affiliation{Artsakh State University,  3 Mkhitar Gosh, Stepanakert, Armenia}

\author{Vahagn Yeghikyan}
\affiliation{Yerevan State University, 1 Alex Manoogian, 375025 Yerevan, Armenia}

 \begin{abstract}
We exclude  the center of mass of the $N$-particle rational Calogero
model and  consider the angular part of the resulting Hamiltonian.
We show that it describes the motion of the particle on
$(N-2)$-dimensional sphere interacting with $N(N-1)/2$ force centers
with Higgs oscillator potential. In the case of four-particle system
these force centers  define the vertexes of an Archimedean solid
called cuboctahedron.
\end{abstract}

%\pacs{PACS numbers 14.80.Hv 02.30.Ik 03.65.-w 11.30.Pb}

%\date{August 4, 2008}

\maketitle

\section{Introduction}
The Calogero model \cite{calogero69,calogero71,moser}
and its various extensions and generalizations
play a distinguished role among other multi-particle integrable
systems. They have attracted much attention  due to their
rich internal structure and numerous applications in many areas of
physics (see, e.g., the recent review \cite{polychronakos} and
references therein).

In the
continuum or thermodynamic limit, i.e. for large particle numbers,
the Calogero model gives rise to a Yang-Mills theory \cite{QCD} on a cylinder,
while its superconformal extension
describes a black hole in the near-horizon limit \cite{black-hole}.
In this limit, the  system have soliton solutions corresponding to the
fundamental excitations \cite{soliton}.

The quantum Calogero model describes free
particles with fractional statistics whose type is determined by the
interaction strength \cite{frac-stat}. Moreover, the
variational ground state  of the fractional quantum Hall effect
(known as the Laughlin state \cite{laughlin}) can be considered as
some deformation of the ground state of Calogero model \cite{Hall}.
The trigonometric analogue of the model \cite{sutherland} is related to integrable
spin-1/2 chains with long range interactions, which possess a
resonating-valence-bond ground state \cite{spin}.
Recently, the relation to the Benjamin-Ono equation
arising in the hydrodynamics of stratified fluids
has been established \cite{hydrodyn}.

The Calogero model and its modifications appear also in matrix models
\cite{matrix-model}, $W_\infty$-algebras \cite{W-alg}, Yangian quantum groups
\cite{yangian}, random matrices \cite{rand-mat}
and many other areas of physics and mathematics.

In this article, we will study the classical rational Calogero
system without confining potential.
It describes one-dimensional particles with inverse-square
interaction \cite{calogero69,calogero71,moser}:
\be
\label{cal} \
{\cal H}=\frac{1}{2}\sum_{i=1}^N{p}^2_i
+\sum_{i<j}\frac{g}{(x_i-x_j)^2},\qquad \{p_i, x_j\}=\delta_{ij}.
\ee
%where $g>0$ determines the interaction strength.
One of the important features of the system is
its manifest conformal invariance, which was essential for the
invention of the model, as well as for its further studies.
%Note that at the quantum level the Hamiltonian \eqref{cal}
%is well defined only for $g>3/4$. For $0<g< 3/4$ it is not self-adjoint
%but admits a one-parameter family of self-adjoint extensions, which
%leads to inequivalent quantizations of the system.
%The corrected Hamiltonian has a bound state with negative energy and breaks
%the scale invariance.

In the pioneering paper \cite{calogero69}, the three-particle model had been
considered first. After excluding the center of mass (with \'a priory
conserving momentum) and taking into account
the conformal invariance, the model was reduced to a one-dimensional
exactly solvable system on  circle considered by Jacobi in the
middle of XIX century \cite{jacobi}:
\be
{\cal I}=
\frac{p_{\varphi}^2}{2}+\frac{9g}{2\cos^23\varphi}. \label{2ph}
\ee
For more particles, the analysis of the Calogero model becomes more
complicated. In particular, the construction of the complete set of
the constants of motion  assumes the use of the powerful method of
Lax pair \cite{moser}. This approach allowed to relate the Calogero
system to $A_{N-1}$ Lie algebras, as well as to construct its
integrable modifications related to other Lie algebras
\cite{algebras}. The  Calogero systems can be obtained from the
free-particle system by an appropriate reduction procedure
known as the projection method \cite{perelomov}. Recently, it has been
generalized to the Calogero model extensions corresponding to the root
systems \cite{sasaki}.

However, the analog of the system (\ref{2ph})
has not been properly studied
for the case of more than three particles. Such a study
would be an interesting problem  from  few  viewpoints.

Already in the pioneering papers \cite{calogero69,calogero71} it was
observed that the spectrum of the Calogero model with additional
oscillator potential is  similar to the spectrum of free
$N$-dimensional oscillator. It was claimed there that a similarity
transformation to the free-oscillator system may exist, at least, in
the part of  Hilbert space. However,  this transformation has been
written explicitly only three decades later \cite{hindu}. In  Ref.
\onlinecite{galajinsky}, it has been  related to  the conformal
group $SU(1,1)$. This similarity transformation  has a very
transparent geometric explanation for the two-particle Calogero
model (the "conformal mechanics"): it corresponds to the inversion
in the Klein model of the Lobachevsky space, which describes the
phase space of the system. A natural way to extend this picture to
the multi-particle Calogero system is to identify the coordinates of
its "radial" part with the coordinates of the Klein model. In other
words, one must extract  and investigate the angular part of the
system.

Another motivation is connected with the superconformal extensions
of Calogero model. In  Ref. \onlinecite{galajinsky}, the authors
suggested to use the aforementioned  similarity transformation for
the construction of ${\cal N}=4$ superconformal  Calogero system
\cite{galaj4}. However, the suggested algebraic  scheme is  quite non-trivial,
and for the cases of four and more  particles  it  assumes the solving of
WDVV and other partial differential equations. On the other hand,  in Ref. \onlinecite{krivonos}
the  superconformal extension of the
 three-particle Calogero model was constructed in a transparent way within the
 superfield approach
by extracting the model (\ref{2ph}) from the initial system. This
approach seems to be applicable to any-particle Calogero system,
under assumption that its  angular part  admits ${\cal N}=4$
supersymmetric extension.

The other stimulation for the study of the angular part of the
Calogero model the translation of the discrete symmetries of the
one-dimensional multi-particle system to the higher-dimensional
one-particle one. This would  provide us with   \'a priori
integrable higher-dimensional one-particle system with some discrete
symmetry.

The purpose of the current article is the investigation of the
angular part of the $N$-particle Calogero model with the excluded
center of mass. Like in the three-particle case, it is a constant of
motion. Hence, its Poisson brackets with the Liouville constants of
motions can generate the additional constants of motions. This can
give a simple explanation of the superintegrability of Calogero
model observed many years ago by Wojciechowski \cite{supint}. We
show  that the angular part itself describes a particle on the
$(N-2)$-dimensional sphere, which interacts with the $N(N-1)/2$
force centers by the Higgs oscillator low. Briefly speaking, we have
a $N(N-1)/2$-center  $(N-2)$-dimensional Higgs oscillator. For the
$N=4$ case  corresponding to particle motion on two-dimensional
sphere, the force centers  are located at  the vertexes of the
Archimedean solid cuboctahedron. This observation opens few horizons
for further study of the Calogero model. In particular, the
investigation of the angular part of the spin-Calogero model and/or
its supersymetric extentions becomes especially important due to the
possibility of  applications in the solid state physics.

The paper is organized as follows.

In Section 2 we exclude the center of mass from the $N$-particle
Calogero model and obtain a $(N-1)$-dimensional system characterized
by  $N(N-1)/2$ unit vectors. These vectors correspond to the
positive roots of the Lie algebra $A_{N-1}$. Then we show that the
angular part of the reduced system describes a particle on
$(N-2)$-dimensional sphere interacting with the vertexes of the
aforementioned vectors by the Higgs oscillator low.

In Section 3 we illustrate the general analysis  of Section 2 for
the simplest nontrivial case of three-particle Calogero model. The
corresponding angular part is given by \eqref{2ph}. We show that its
Poisson bracket with the third-order Liouville constant of motion
coincides with the additional constant of motion. Then we derive an
explicit functional dependence between four constants of motion
(including the Hamiltonian).

In Section 4 we consider the four-particle system. We show that its angular part
describes a multicenter Higgs oscillator on two-dimensional sphere with force centers
located at the vertexes of the cuboctahedron.

In the last section we summarize the results and discuss the possible extension
to the supersymmetric  Calogero systems.

\section{Center-of-mass system}
In order to decouple the center of mass, we introduce
a new coordinate system for $N$-particle Calogero model.
\begin{equation}
y_0=\frac{1}{\sqrt{N}}\sum _{i=1}^N x_i,\qquad
y_k=
\frac{1}{\sqrt{N-k+1}}
\left(
\sqrt{N-k} \ x_{k}-\frac{1}{\sqrt{N-k}}\sum _{i=k+1}^N x_i
\right),
\quad 1\le k\le N-1
\label{trans}
\end{equation}
Here $y_0$ describes the center-of-mass movement, while the remaining $y_k$
describe the motion in the certer-of-mass system.
The transformation \eqref{trans} is chosen to be orthogonal:
$$
\sum_{k=1}^N (dx_k)^2= (dy_0)^2+\sum_{k=1}^{N-1} (dy_k)^2.
$$
Hence,  the inverse transformation $x_k=\sum_{n=1}^N A_{kn}y_n$ coincides with its
transpose:
$$
A_{km}=
\begin{cases}
1/\sqrt{N}& \text{for $m=0$}\\
-1/\sqrt{(N-m+1)(N-m)}& \text{for  $k>m\ge1$}\\
\sqrt{N-k}/\sqrt{N-k+1}& \text{for $m=k$}\\
0& \text{for other $m$}
\end{cases}
$$
Using these formulae, we rewrite the Hamiltonian of the Calogero
model in terms of the center-of-mass variables:
$$
{\cal H}=\frac{1}{2}\sum_{i=1}^N{p}^2_i
+\sum_{i<j}\frac{g}{(x_i-x_j)^2}=\frac{p^2_0}{2}+{\cal \tilde H}.
$$
Here $p_0=\sum_{i=1}^N {p}_i$ is the conserved total momentum of the
$N$-particle Calogero system. The last  term is given by the
expression
\be
{\cal \tilde H}=\frac{1}{2}\sum_{i=1}^{N-1}{p}^2_i+
\sum_{a=1}^{N(N-1)/2}\frac{g}{2\left(\sum_{k=1}^{N-1}
b^a_ky_k\right)^2},\qquad \{p_i, y_j\}=\delta_{ij},
\label{n-1}
\ee
where $a\equiv(i,j)$  enumerates the
pairs of interacting particles, $p_i$ (we keep the old notation for them)
are the new momenta conjugated to $y_i$,   and
\be
\label{b}
b^{a}_k=b^{ij}_k=\frac{A_{ik}-A_{jk}}{\sqrt{2}}.
\ee
It is clear that the  Hamiltonian with the excluded center of mass
${\cal \tilde H}$ is a constant of motion of the original system \eqref{cal}.
From the orthogonality of the matrix
$A_{ik}$ we have:
\be
\label{scalar}
\sum_k (b^{ij}_k)^2=1, \qquad
\cos\alpha_{ij,i'j'}= \sum_k b^{ij}_kb^{i'j'}_k=
\frac{1}{2}\sum_k(A_{ik}-A_{jk})(A_{i'k}-A_{j'k})=\frac{1}{2}(\delta_{ii'}+\delta_{jj'}-\delta_{ij'}-\delta_{i'j}).
\ee
So, the coordinates ${\bf b}^a=(b^{a}_1,\ldots, b^{a}_{N-1})$
define unit vectors in $(N-1)$-dimensional space, while $\alpha_{ij,i'j'}$
are the angles between them.
In fact, ${\bf b}^a$ correspond to the positive
roots of the Lie algebra $A_{N-1}$ (rescaled by the factor $1/\sqrt{2}$).
Indeed, the potential of original model \eqref{cal} can be presented as the inverse-square
sum over all positive roots $\Delta_+$ of $A_{N-1}$ multiplied by the particle coordinates:
$\sum_{\mathbf{\alpha}\in\Delta_+}g/(\mathbf{\alpha}\cdot \mathbf{x})^2$ \cite{perelomov}.
The orthogonal transformation \eqref{trans} acts on those roots by mapping them
onto ${\bf b}^a$.
Therefore, the last vectors  define the same root system.

The reduced system \eqref{n-1}, \eqref{b} can be
interpreted as a one-particle system in $(N-1)$-dimensional space.
Let us  extract the radius $r$ of the obtained system.
This can be done, for instance, in hyperspherical coordinates,
where the Hamiltonian \eqref{n-1} takes the following form:
\be
%{\cal H}=\frac{p^2_0}{2}+{\cal \tilde H},
%\qquad
{\cal \tilde H}=\frac{p^2_r}{2}+\frac{{\cal I}(p_{\varphi_\alpha},\varphi_\alpha)}{r^2},
\qquad
{\cal I}(p_{\varphi_\alpha},\varphi_\alpha)=\frac{K_\text{sph}(p_{\varphi_\alpha},\varphi_\alpha)}{2}+\sum_a\frac{g}{{2\cos^2\theta_a }},
\qquad
\{p_{\varphi_\alpha}, \varphi_\alpha\}=\delta_{\alpha\beta},
\ee
where $\alpha,\beta=2,\ldots,N-1$.
Here  $K_\text{sph}$ is the standard kinetic term of the
particle on the hypersphere $S^{N-2}$ with unit radius,
$\theta_a$ is the  angle between  ${\bf b}_a$ and the unit vector
${\bf n}={\bf r}/r$ directed from the hypersphere center  to the particle.

Since ${\cal I}$ is independent from the radial coordinates
$p_r$ and $r$, it commutes with the Hamiltonian ${\cal \tilde H}$. {\sl So, it
is a constant of motion of the Calogero model.} Note that this
integral is quadratic on the momenta (while in the standard Lax pair
approach, the only constant of motion, which is quadratic on
momenta, is the Hamiltonian). It is easy to verify that
any other integral being in involution with ${\cal I}$
must depend on the radial coordinated through  the Hamiltonian ${ \cal \tilde H}$.
Therefore, it is not in involution, at least, with the integrals  having an odd
order on $p_r$ and is functionally independent
from  the $N$ commuting integrals constructed using the Lax pair.
The matter is that the Calogero model is not only an integrable
 in the Liouville sense (which means the existence of $N$
integrals being in involution) but is maximally
superintegrable  with $2N-1$ functionally independent integrals
\cite{supint} (see also \cite{supinto}).
Clearly, ${\cal I}$ is a function of this complete set of the constants of motion.
Moreover, its Poisson bracket action on the Liouville
constants of motions generates the additional constants of motion,
which are responsible for the superintegrability.
% Thought the additional constants of motion have  higher (than two) order on momenta.

It can be considered  as the Hamiltonian of a particle moving on
the $(N-2)$-dimensional sphere with $N(N-1)/2$ force centers defined
by the vectors ${\bf b}^a$. Since this system is invariant under
reflections ${\bf b}^a\to -{\bf b}^a$ for any $a$, sometimes it is
reasonable to consider the $N(N-1)$  properly located force centers.
In order to clarify the physical meaning of the obtained  system,
let us rewrite its potential as follows
$$
V_{\rm sph}=\sum_a\frac{g}{2\cos^2\theta_a}=\frac{N(N-1)g}{4}+\frac{g}{2}\sum_a\tan^2\theta_a\;.
$$
Recall now that
$$
V_{\rm Higgs}=\frac{\omega^2r^2_0\tan^2\theta}{2}
$$
is the well-known  potential of the Higgs oscillator.
It generalizes the ordinary oscillator potential for the sphere with
the radius $r_0$ and inherits all its hidden symmetries  \cite{higgs}.

{\sl
Hence, we obtained the integrable  $N(N-1)/2$-center   $N$-dimensional Higgs oscillator of the frequency
$\omega=\sqrt{g}$.
}

The location of the force  centers is quite rigid and deserves to be considered in more details.
Note that the Higgs oscillator has been invented about thirty years ago
and has been studied the hundreds of papers so far (see, e.g. \cite{hosc} and refs therein).
Nevertheless, its anisotropic version was found quite recently \cite{anosc}, whereas
the two-center version is not known yet, up to our knowledge.

\section{Three-particle case: circle}

\begin{figure}[t]
\begin{minipage}[t]{0.4\textwidth}
\includegraphics[height=0.8\textwidth]{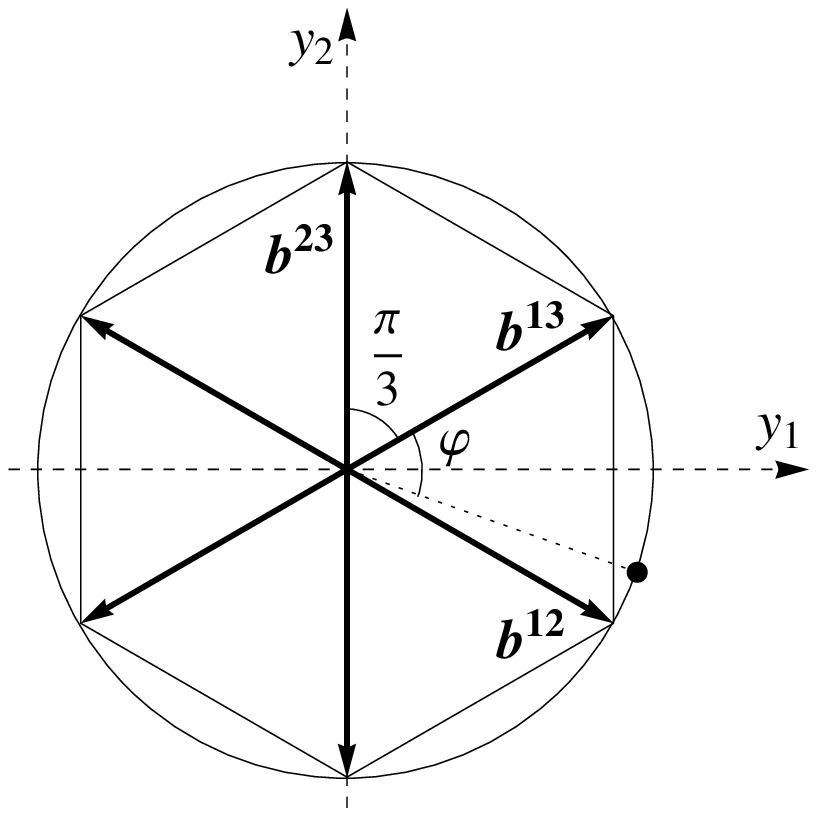}
\caption{\label{fig:1}
 The force centers (${\bf b}^{12}$, ${\bf b}^{23}$, ${\bf b}^{13}$ and their
 opposites), which form the root system of $su(3)$ and constitute an hexagon.
 The angle $\varphi$ describes the position of a particle on cycle.
 }
\end{minipage}
\hspace{0.1\textwidth}
\begin{minipage}[t]{0.4\textwidth}
\includegraphics[width=0.9\textwidth]{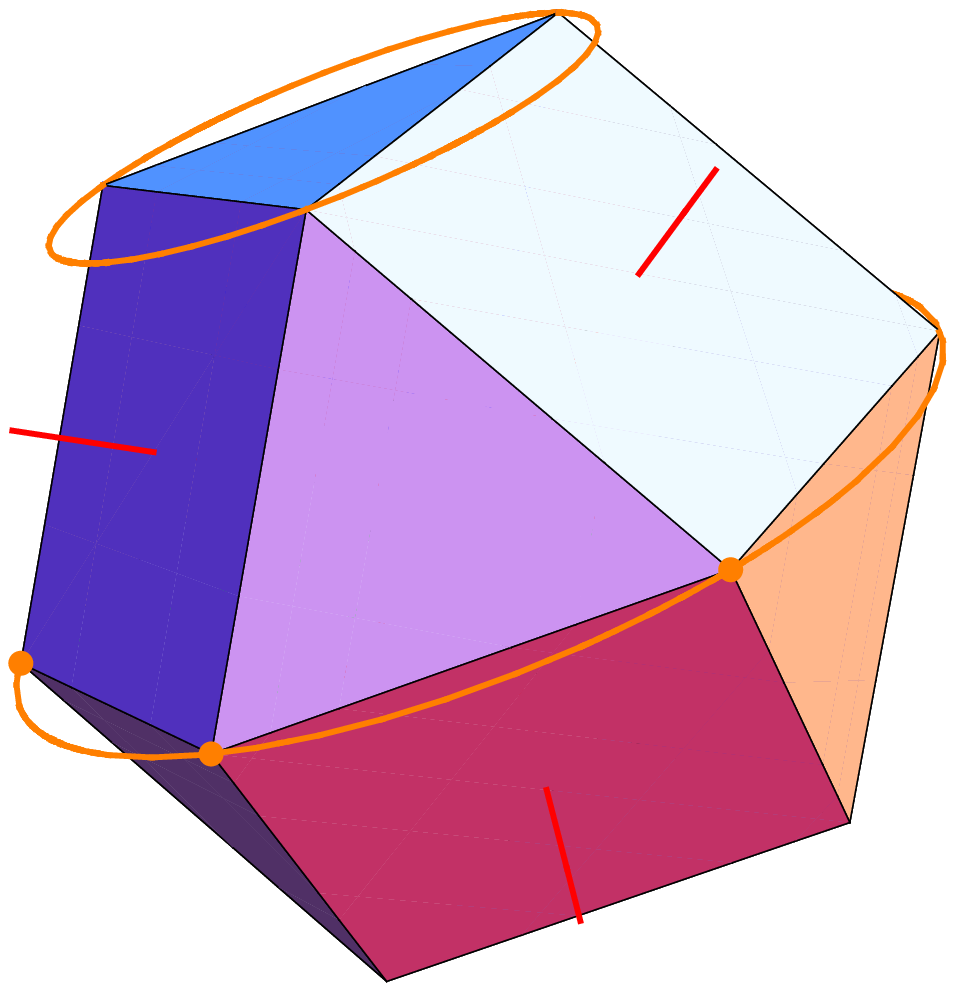}
\caption{
\label{fig:2}
The vectors \eqref{cuboc} together with their opposites
form a cuboctahedron and are equivalent to the root system of $su(4)$.
The bold points on the large cycle correspond to
${\bf b}^{23}$, ${\bf b}^{34}$ and ${\bf b}^{24}$ while
the small cycle contains the vertexes of the remaining three vectors.
The bold lines are the axes of the coordinate system \eqref{a-system}.
}
\end{minipage}
\end{figure}

The simplest system  is the angular part of three-particle model
considered in the pioneering paper by Calogero \cite{calogero69}.
Actually, this system was considered in the middle of XIX by Jacobi
\cite{jacobi} (see also \cite{perelomovb}). For $N=3$, we get a
particle on  circle $S^1$ with three force centers defined by the
unit vectors ${\bf b}^{12}$, ${\bf b}^{23}$ and ${\bf b}^{13}$. The
angles between them are equal to $\pi/3$ and $2\pi/3$ (Fig.
\ref{fig:1}):
$$
\cos\alpha_{12,13}=\cos{\alpha}_{13,23}=1/2, \qquad
\cos\alpha_{12,23}=-1/2.
$$
The above vectors make up the set of positive roots of $A_2\equiv
su(3)$ Lie algebra. Completing them by the oppositely directed
vectors  corresponding to the negative roots, we obtain a system
with six force centers. The  angular part of the Hamiltonian \be
\label{I} {\cal
I}=\frac{p_{\varphi}^2}{2}+\frac{g}{2\cos^2\varphi}+\frac{g}{2\cos^2(\varphi+\pi/3)}+
\frac{g}{2\cos^2(\varphi-\pi/3)}=
\frac{p_{\varphi}^2}{2}+\frac{9g}{1+\cos 6\varphi} \ee coincides
with \eqref{2ph}. It is invariant under the rotation on $\pi /3$ and
the reflection $\varphi\to-\varphi$, which generate the symmetry
group $D_6\equiv S_3\otimes Z_2$  of the hexagon (Fig. \ref{fig:1}).
Here $S_3$ is the symmetric group of three-particle permutations,
which ${\cal I}$ inherits from the original Calogero Hamiltonian
\eqref{cal}. The $Z_2$-symmetry corresponds to the
reflection-invariance $x_i\to-x_i$ of \eqref{cal}. The integrability
of this system is obvious. Note that the splitting of the
three-particle Calogero Hamiltonian on the angular and radial parts
has been used in Ref. \onlinecite{feher} for the detailed study of
the quantization.

Let us briefly discuss the relation of the system on circle with the
superintegrability of three-particle Calogero model. Note that its
superintegrability  was studied  in detail (see \cite{sup3} and
refs. therein). In the center-of-mass system, three from the five functionally independent
constants of motion of the original Calogero system survive. Namely,
the Hamiltoinian of the two-particle system
$
{\cal \tilde H}=p_r^2/2+{\cal I}/{r^2}
$
and its constant of motion
\be
\label{F}
{\cal F}=
   \left(p_r^2 - \frac{6 {\cal I}}{r^2}\right)  p_r \sin 3 \varphi +
   \left(3p_r^2 - \frac{2 {\cal I}}{r^2}\right)  \frac{p_\varphi \cos 3\varphi}{r}
\ee
are reduced from the second and third order (on momentum) Liouville constants of motion.
Similarly, the third conserved quantity
\be
\label{K}
{\cal K}=
     \left(p_r^2 - \frac{6{\cal I}}{r^2} \right) p_r p_\varphi \cos 3\varphi
     -  \left(3p_r^2 - \frac{2{\cal I}}{r^2} \right)\frac{2{\cal I}\sin 3 \varphi}{r}
\ee is inherited from the additional third order constant of motion
of three-particle Calogero system. The integrals ${\cal \tilde H}$,
${\cal F}$, and ${\cal K}$ are functionally independent. We have
expressed them in terms of the angular part of the Hamiltonian
\eqref{I}, which also conserves. Its Poisson bracket action maps the
Liouville integral to the additional one: \be \label{adI} \{{\cal
I,F}\}=3 {\cal K}, \qquad \{{\cal I,K}\}=-6 {\cal I F}. \ee The four
quantities ${\cal\tilde H}$, ${\cal I}$, ${\cal F}$, and ${\cal K}$
form an overcompleted set of constants of motion. They and subjected
to the algebraic relation \be \label{Ksq}
 {\cal K}^2 + 2{\cal I F}^2= 8{\cal\tilde H}^3(2{\cal I}-9g),
 \qquad \text{or} \qquad
  {\cal I}=\frac{{\cal K}^2+72g{\cal \tilde H}^3}{16{\cal\tilde H}^3-2{\cal F}^2}.
\ee
Hence, one can choose ${\cal\tilde H}$, ${\cal I}$, and ${\cal F}$ as a complete set of
functionally independent conserved quantities. The first two of them are quadratic on
momenta, which ensures the separation of variables in the system.

It is easy to verify that the Poisson brackets \eqref{adI} are
in consistency with the relation \eqref{Ksq}.
Finally, using \eqref{Ksq} and the first equation in \eqref{adI}, we obtain the
Poisson bracket between two third-order integrals:
$$
\{{\cal K,F}\}=3(8{\cal \tilde H}^3- {\cal F}^2)=3\frac{{\cal K}^2+9g{\cal F}^2}{2{\cal I}-9g}.
$$

%But this new constant of motion is precisely the fourth  constant of motion
%constructed within Lax pair approach!

%\setcounter{equation}{0}
\section{Four-particle system: sphere}
In the four-particle case, everything becomes much more complicated.
In the same way, we obtain a system on the sphere with {\em six } force centers
defined by the unit vectors ${\bf b}^a$  with the following Cartesian coordinates of the
ambient $\DR^3$ space:

\be
\label{cuboc}
\begin{aligned}
&{\bf b}^{12}=\left(\sqrt{2\over 3},\ -\frac{1}{\sqrt{3}},\ 0 \right),
\qquad &&
{\bf b}^{13}=\left(\sqrt{2\over3},\ \frac{1}{2\sqrt{3}},\ -\frac{1}{2}\right),
\qquad &&
{\bf b}^{14} = \left(\sqrt{2\over3}, \ \frac{1}{2\sqrt{3}}, \ \frac{1}{2}\right),
\\
&{\bf b}^{23}=\left(0,\ {\sqrt{3}\over2},\ -\frac{1}{2}\right),
\qquad &&
{\bf b}^{24}=\left(0,\ {\sqrt{3}\over2},\ \frac{1}{2}\right),
\qquad &&
{\bf b}^{34}=(0, \ 0, \ 1).
\end{aligned}
\ee
The vertexes of ${\bf b}^{ij}$ and their opposite vectors  form an Archimedean solid called
\textbf{\emph{cuboctahedron}} (Fig.\ref{fig:2}).
This polyhedron, like cube, has the octahedral symmetry  $O_h\equiv S_4\otimes Z_2$ of order 48.
%which reflects the discrete symmetries of the original four-particle Calogero system.
Here $S_4$ is the symmetric group of four-particle permutations, which preserve the
original Calogero Hamiltonian \eqref{cal}.
Note that $S_4$ is isomorphic to the Weyl group of $A_3$ Lie algebra and preserves
the orientation of cuboctahedron. The $Z_2$ symmetry corresponds to the reflection
$x_i\to-x_i$ of all four coordinates. In Lie algebraic description, it corresponds to
the reflection symmetry of $A_3$ Dynkin diagram.

Note that the vectors ${\bf b}^{23}$, ${\bf b}^{34}$ and ${\bf b}^{24}$
belong to the "equatorial" plane, the angles between them are equal to $\pi/3$ and $2\pi/3$.
Their vertexes and the vertexes of the opposite vectors form an hexagon (Fig. \ref{fig:2}).
This is precisely the same picture as in the three-particle Calogero model (see Fig. \ref{fig:1}).
The endpoints of  the vectors ${\bf b}^{12}$, ${\bf b}^{13}$, ${\bf b}^{14}$ are located  on a plane
parallel to the equatorial one (Fig. \ref{fig:2}). The distance between both planes is $\sqrt{2/3}$.
They form the (regular) triangular face of the cuboctahedron, which is shifted by the angle $\pi/6$ with respect
to the triangle $({\bf b}^{23},{\bf b}^{34},-{\bf b}^{24})$.

Let us choose Cartesian coordinates with  the first axis directed
along ${\bf b}^{13}$ while the second one belonging to the plane
formed by ${\bf b}^{12}$ and ${\bf b}^{13}$. The frame directions
then are orthogonal to the triangles of the cuboctahedron (Fig.
\ref{fig:2}). In the respective  spherical coordinates, the angular
part of the Hamiltonian reads: \be {\cal
I}=\frac{p^2_\theta}{2}+\frac{p^2_\varphi}{2\sin^2\theta}+
\frac{9g(8-\tan^2\theta)^2}{2(3\tan^2\theta-8+\tan^3\theta\cos3\varphi)^2}
+\frac{12g}{3\tan^2\theta-8+\tan^3\theta\cos3\varphi}+\frac{9g}{4\sin^2\theta
(1+\cos 6\varphi )}. \label{s23}
\end{equation}
The invariance under $Z_3$ group of the rotation on $2\pi/3$ along the third axis is apparent.

The potential \eqref{s23} is really horrible.
It is difficult to believe, that the system with such potential could be integrable,
or could admit a separation of variables.
However, the Hamiltonian  can  be represented in a much simpler form. Indeed,
there are  three pairs  of the orthogonal vectors
$
{\bf b}^{12}\cdot{\bf b}^{34}={\bf b}^{13}\cdot{\bf
b}^{24}={\bf b}^{14}\cdot {\bf b}^{23}=0
$.
Taking the vector products of these pairs, one can find out that they  form an orthogonal
frame:
\be
\label{a-system}
{\bf a}_1\equiv {\bf b}^{12}\times{\bf b}^{34},
\qquad
{\bf a}_2\equiv {\bf b}^{13}\times{\bf b}^{24},
\qquad
{\bf a}_3\equiv
{\bf b}^{14}\times{\bf b}^{23}\, :
\qquad
{\bf a}_i\cdot{\bf a}_j=\delta_{ij}.
\ee
The  vectors ${\bf a}_i$ are normal to the squares of the cuboctahedron (Fig.\ref{fig:2}).
In this coordinate system, the  Hamiltonian (\ref{n-1}) looks like
\be
\label{H-u}
{\cal \tilde H}=\sum_{i=1}^3\frac{p^2_i}{2}+
\sum_{1\le i<j\le3}\left(\frac{g}{(u_i-u_j)^2}+ \frac{g}{(u_i+u_j)^2}\right),\qquad \{p_i,u_j\}=\delta_{ij},
\ee
where, again, we keep the old notations for the new momenta.
This is the three-particle $D_3$ Calogero model \cite{perelomov}.
However, this is an expected result,
since the  diagrams $D_3$ and $A_3$ coincide and define the same algebra
(in the Dynkin classification, $D_n$ is defined for $n\ge4$).

The  angular part of this Hamiltonian has the following form: \be
{\cal I}=\frac{p^2_\theta}{2}+\frac{p^2_\varphi}{\sin^2\theta}+
\frac{4g}{\sin^2\theta} \left[\frac{1}{1+\cos
4\varphi}+\frac{k-6}{k-8+8/k -k\cos 4\varphi} +
\frac{4(k-16+16/k)}{\left(k-8+8/k -k\cos 4\varphi\right)^2}\right],
\label{z4} \ee where
$$
k=\tan^2\theta=\frac{1-\cos 2\theta}{1+\cos 2\theta}.
$$
In these new spherical coordinates, the invariance under $Z_4$ rotations
$\varphi\to\varphi+\pi/4$ and spatial reflections $\theta\to \pi-\theta$,
which are a subgroup in $O_h$,  is transparent.

As was explained above and showed explicitly for three-particle
system, ${\cal I}$ can be expressed in terms of the five integrals
of the maximally superintegrable Hamiltonian \eqref{H-u}. It seems
that the two additional integrals of ${\cal\tilde H}$ can be
obtained from the Liouville integrals by Poisson bracket action with
${\cal I}$ like in  the three-particle case \eqref{adI}.

Since the spherical system \eqref{z4} was obtained from the Calogero model,
it is also integrable. Its constants of motion can be obtained from those of the original model.

\section{Summary and discussion}

In conclusion, let us emphasize the main results of the current
article.

We have found that the angular part of $N$-particle rational
Calogero model (in the center-of-mass system) gives rise to the
$N(N-1)/2$-center hyperspherical (Higgs) oscillator. Its relations
with the superintegrability of the Calogero system has been
discussed briefly.

For the three-particle system, we have related the angular part of
the Calogero Hamiltonian (which is quadratic on momentum) with the
higher order constants of motion.

For the four-particle Calogero model, the force centers are located
at the vertexes of the Archimedean solid cuboctahedron.

We are planning in future studies to investigate the reflection of
the superintegrability of the rational four-particle Calogero model
on the properties of corresponding spherical system and, presumably,
to extend this investigation to the  case of $N>4$ particles. In
particular, this would help us to answer whether the the Hamiltonian
(\ref{z4}) admits a separation of variables. We expect that the
Poisson bracket of the angular part with the Liouville constants of
motion will generate the additional serie of conserved quantities,
which are responsible for the superintegrability.

Another  task is to extract and study
the angular parts of Calogero models associated with other Lie
algebras. It is obvious that they are also connected with a
multi-center spherical integrable system related with
(high-dimensional) polyhedra.

It is clear that the proposed construction is applicable for the
quantum spin-Calogero model too. In that case the force centers of
the angular part will be equipped by the spin-1/2 exchange
interactions \cite{poly93-1}. In this context, the relation of the
four-particle system with the cuboctahedron becomes important. We
expect that using the freezing trick \cite{poly93-2}, one can obtain
an integrable spin lattice system on cuboctahedron. Note that
recently the Heisenberg spin systems on large magnetic molecules
has been investigated intensively \cite{molec-mag}. In particular,
cuboctahedric molecular magnets with nearest-neighbor Heisenberg
interaction has been obtained experimentally and studied
theoretically \cite{cuboct}. It will be interesting to consider from
this viewpoint $N=2$ supersymmetric Calogero system \cite{freed90}.
The construction of the ${\cal N}=4$ supersymmetric counterpart of
the suggested  system and the study of its integrability is also
important. For the cuboctahedric system, the solution of this task
is more or less obvious: we just need to check  whether is it
possible to represent the Hamiltonian (\ref{s23}) or (\ref{z4}) in
the form \cite{surprize}
$$
{\cal I}=
\frac{1}{g(z,\bar z )}\left(\pi\bar\pi+\frac{F(z){\overline F}(\bar z )}{(1+\lambda(z){\overline\lambda}(\bar z))^2}\right) ,
\quad\{\pi, z\}=1,
\qquad {\rm where}\quad  {\overline \lambda}'F= -\lambda{\overline F}.
$$
In this respect, the recent paper \cite{fed},  where
the supersymmetric Calogero model was related
with the nontrivial deformation of spin-Calogero model, deserves to be mentioned.

\begin{acknowledgments}
 We are grateful to Sergey Krivonos for
the discussions on three-particle Calogero system, which prompted
us to this study, and to  Vadim Ohanyan and Olaf Lechtenfeld for the interest in work and useful comments.
The work was supported by grants NFSAT-CRDF UC-06/07, INTAS-05-7928 (T.H. \& A.N.), ANSEF-1386PS (T.H.),
ANSEF-1730PS (A.N. \& V.Y.), and by the Artsakh Ministry of Education and Science (A.N. \& T.H.).
\end{acknowledgments}

\end{document}